\documentclass[11pt]{article}

\usepackage{amssymb}
\usepackage[pdftex]{color,graphicx}
\usepackage[paperwidth=5.3in, paperheight=7.5in, top=.55in, bottom=.5in, left=.14in, right=.14in]{geometry}
\usepackage{textcomp}

\newcommand{\bmax}{b_\textsl{max}}

\newcommand{\ipar}{\begin{list}{\hbox{$\:$}}{} \item[] }
\newcommand{\rapi}{\end{list}}

\begin{document}

\title{\bf The Bounded Edge Coloring Problem \\
and Offline Crossbar Scheduling}
 
\author{Jonathan Turner}
\date{\normalsize WUCSE-2015-07}
\maketitle

\begin{abstract}
This paper introduces a variant of the classical edge coloring problem
in graphs that can be applied to an offline scheduling problem for crossbar switches.
We show that the problem is {\sl NP}-complete, develop three lower bounds
bounds on the optimal solution value and evaluate the performance 
of several approximation algorithms, both analytically and experimentally.
We show how to approximate an optimal solution with a worst-case performance ratio of $3/2$ and
our experimental results demonstrate that the best algorithms produce results that very closely 
track a lower bound.
\end{abstract}

\pagestyle{plain}

\section{Introduction}
\begin{figure}[h]
\centerline{\includegraphics[width=1.75in]{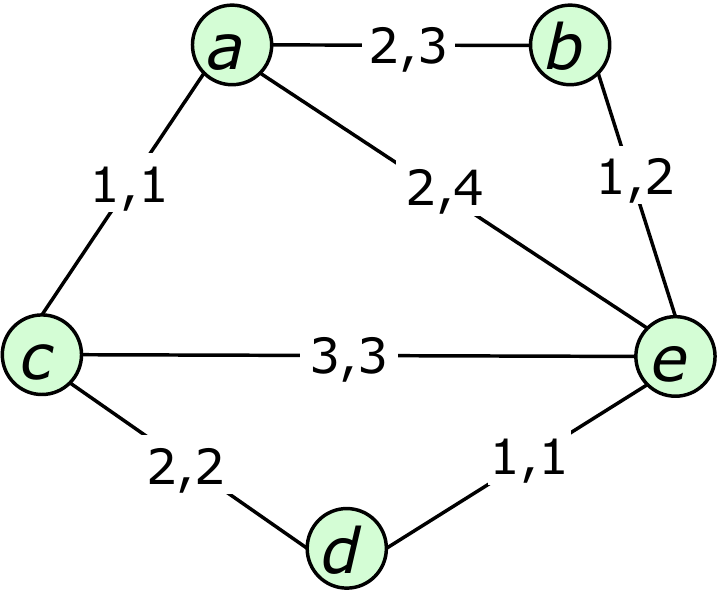}}
\caption{Example of bounded edge coloring}
\label{example1}
\end{figure}

An instance of the {\sl bounded edge coloring problem} is an undirected graph
$G=(V,E)$ with a positive integer {\sl bound} $b(e)$ for each edge $e$.
A {\sl bounded edge coloring} is a function $c$, from the edges to the positive integers,
with $c(e)\geq b(e)$ for all edges $e$ and $c(e_1) \neq c(e_2)$ for all edge
pairs $e_1$ and $e_2$ that have an endpoint in common.
The objective of the problem is to find a coloring in which the largest color
is as small as possible.
An example is shown in Figure~\ref{example1};
the first number labelling each edge is its bound, while the second is
a valid color.

\begin{figure}[h]
\centerline{\includegraphics[width=3.5in]{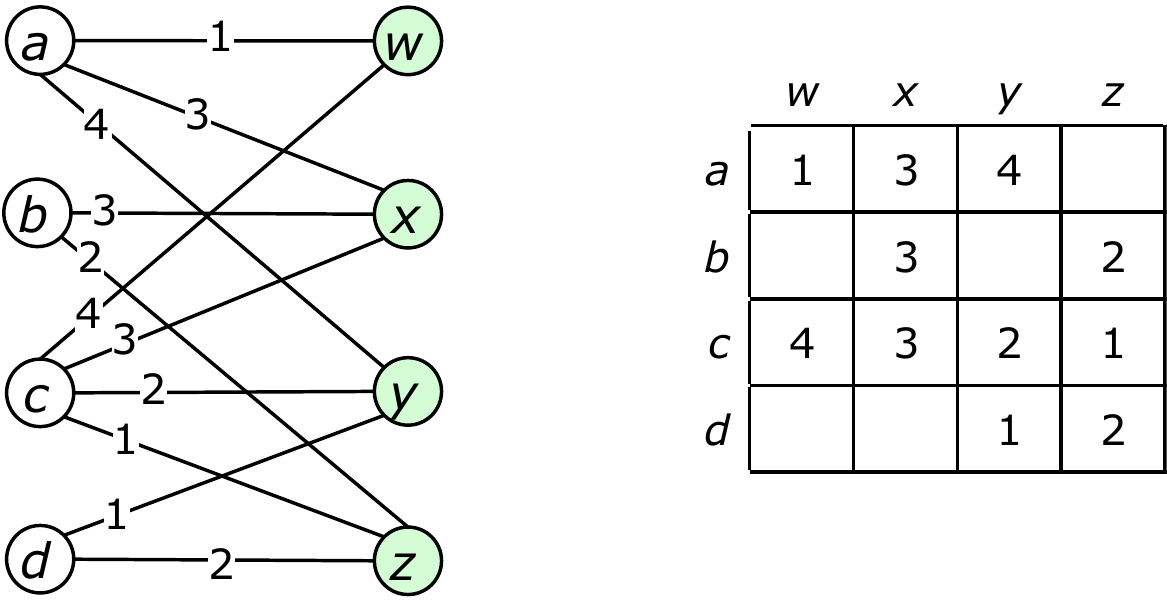}}
\caption{Bipartite instance that satisfies the unique input bounds condition}
\label{example2}
\end{figure}

In this paper, we focus on a restricted version of the problem
in which the graph is bipartite, with the vertices divided between
{\sl inputs} and {\sl outputs}. We also are usually interested in graphs
where the bounds on the edges incident to each input are unique;
we refer to this as the {\sl unique input bounds condition}.
An example of a graph that satisfies the unique input bounds
condition is shown in Figure~\ref{example2}.
The right side of the figure shows a tabular representation of the graph,
where each row corresponds to an input, each column corresponds to
an output and each integer denotes an edge and its bound.
Note that this graph requires colors $[1\ldots 5]$.

The bounded edge coloring problem is an abstraction of an offline version
of the crossbar scheduling problem. In this application, the graph's vertices represent
the inputs and outputs of a crossbar switch, while the edges represent packets to be transferred
from inputs to outputs. The edge bounds represent the arrival times of the packets
and the colors represent the times at which packets are transferred from inputs to
outputs. Since an input can only receive one packet at each time step,
the inputs naturally satisfy the unique input bounds condition.
The objective of the problem is to transfer all packets across the crossbar
in the smallest possible amount of time.

The crossbar scheduling problem has been studied extensively in the online context,
using several distinct performance criteria. For so-called {\sl input queued} switches,
the focus has been on ensuring bounded waiting times in systems subjected to random input traffic.
Many scheduling algorithms have been shown to meet this 
objective~\cite{leonardi-01,mckeown-99a, mckeown-99b}.
More compelling, worst-case results have been shown for {\sl combined input and output queued}
switches, in which the crossbar is capable of transferring packets somewhat faster than they
can arrive at the inputs, or be transmitted from the outputs.
Some scheduling algorithms can match the performance of an idealized {\sl output-queued}
switch when the crossbar is twice as fast as the inputs and 
outputs~\cite{attiya-06,chuang-99,krishna-99}.
 
In section 2, we show that the bounded edge coloring problem is {\sl NP}-complete.
In section 3 we derive several lower bounds on the number of colors required,
and introduce a class of graphs which require substantially more colors than
implied by the weaker of our lower bounds.
In section 4, we describe several algorithms and establish worst-case performance
ratios for some.
Experimental performance results appear in section 5.

\section{Complexity of Bounded Edge Group Coloring}
For the ordinary edge coloring problem, 
we can color any bipartite graph with colors $[1\ldots \Delta]$, where $
\Delta$ is the maximum vertex degree~\cite{BM76}.
Unfortunately, the bounded edge coloring problem is {\sl NP}-complete.
We can show this, using a reduction from the {\sl partial edge coloring completion problem}
for bipartite graphs. In this problem, we are given a bipartite graph with some of its
edges colored, and are asked to complete the coloring using no color larger than
a given integer $k$.
This is a generalization of the {\sl partial latin squares completion problem}
shown to be {\sl NP}-complete in~\cite{CO84}.

\begin{figure}[t]
\centerline{\includegraphics[width=4in]{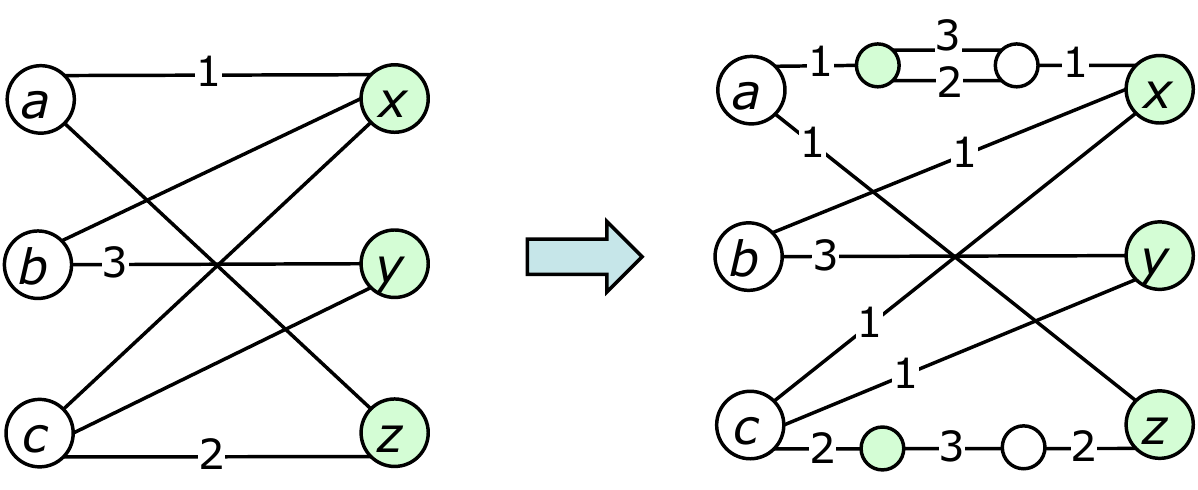}}
\caption{Reduction from edge coloring completion to bounded edge coloring}
\label{npReduction1}
\end{figure}

Let $G=(V,E)$ be a bipartite graph, with partial coloring $c$, using colors $1,\ldots ,k$.
We construct a second bipartite graph $H=(W,F)$ where $W$ includes all vertices in $V$,
plus some additional vertices to be specified shortly. Each uncolored edge in $E$ is included
in $F$ and is assigned a lower bound of 1.
Each edge in $E$ that is colored $k$ is included in $F$ and assigned a lower bound of $k$.
For each edge in $E$ that is colored $k-1$, we include a chain of three edges,
with the inner edge assigned a lower bound of $k$, while the outer two edges are assigned
lower bounds of $k-1$.
For each edge in $E$ that is colored $k-i$, for $i>1$, we include a similar component
with two additional vertices, two ``outer'' edges and $i$ parallel edges joining the two added
vertices.
The outer edges are assigned lower bounds of $k-i$, 
while the inner edges are assigned lower bounds of $k-i+1,\ldots,k$.
The construction of the chains guarantees that in any valid $k$-coloring of $H$, 
the outer edges are assigned the same color that the corresponding
edge was assigned in $G$.
The uncolored edges in $G$ are free to use any color in $1,\ldots,k$.
This construction is illustrated in Figure~\ref{npReduction1} for
the case of $k=3$.
It's straightforward to show that the partial coloring of $G$ can be completed using
colors $1,\ldots,k$ if and only if $H$ can be colored using colors $1,\ldots,k$.

\begin{figure}[h]
\centerline{\includegraphics[width=4.75in]{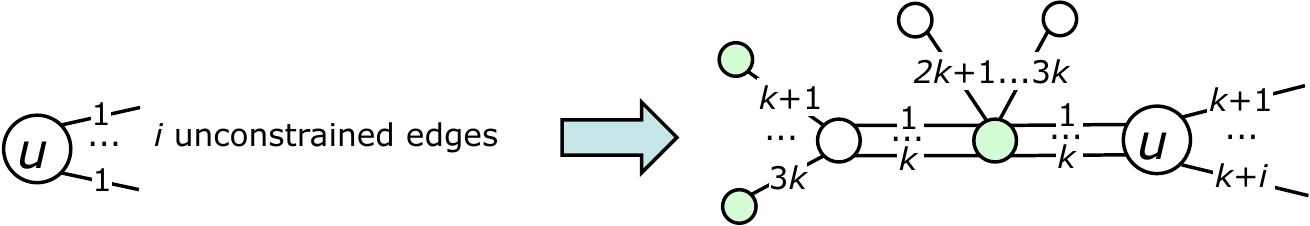}}
\caption{Additional component for each input}
\label{npReduction2}
\end{figure}

\begin{figure}[t]
\centerline{\includegraphics[width=3in]{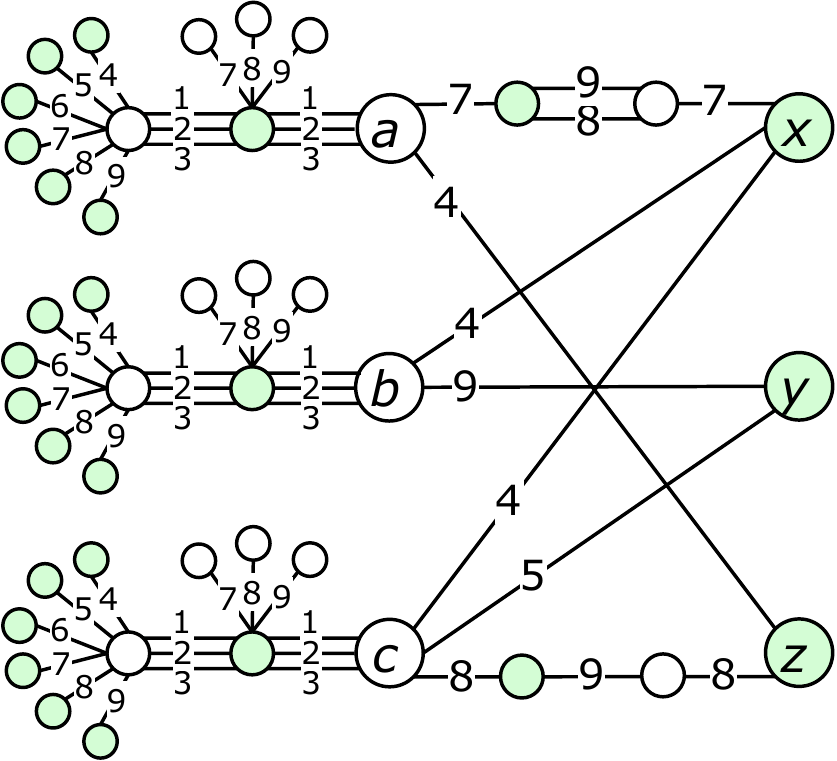}}
\caption{Extension of example in Figure~\ref{npReduction1}}
\label{npReduction3}
\end{figure}

Unfortunately, while this reduction does show that bounded edge coloring is {\sl NP}-complete,
the graph $H$ does not satisfy the unique input bounds condition. To show that instances
that satisfy this condition are also hard to solve, we need to use a more elaborate
reduction. The new reduction starts with the graph $H$ constructed above, and then
adds $2k$ to the lower bounds specified earlier for the components constructed to
handle the pre-colored edges in the original partial edge coloring instance. 
For edges that are uncolored in $G$ and incident to
an input $u$, the corresponding edges in $H$ are assigned distinct lower bounds in
the range $k+1,\ldots,2k$. We then attach an additional component to $u$ that
has the effect of forcing these edges to have colors larger than $2k$.
This construction is illustrated in Figure~\ref{npReduction2}.
This component has two groups of parallel edges joining a pair of vertices to
the original input $u$; these edge groups are each assigned lower bounds of $1,\ldots,k$.
There are two other groups joining additional leaf vertices to the first two vertices. 
One group of ``leaf edges'' has lower bounds of $k+1,\ldots,3k$ and must be
assigned colors equal to their bounds in any valid coloring with a maximum color of $3k$. 
The other has bounds $2k+1,\ldots,3k$ and must also be assigned colors equal to their bounds
in any valid coloring with a maximum color of $3k$. This means that the $k$ parallel
edges incident to $u$ must be assigned colors $k+1,\ldots,2k$ and this is in turn forces
the edges that were uncolored in $G$ to have colors larger than $2k$ in the modified
version of $H$. Figure~\ref{npReduction3} shows the final version of $H$, for the example
in Figure~\ref{npReduction1}.
Observe that all input vertices in the final version of $H$ do in fact have unique lower bounds.
It's straightforward to show that the final version of $H$
can be colored using colors $1,\ldots,3k$ if and only if $G$'s coloring can be completed
using colors $1,\ldots,k$.

\section{Lower Bounds}

In this section, we present several methods to compute lower bounds on the
maximum edge color used by an instance of the bounded edge-coloring problem. 
The first is referred to as the {\sl degree bound}.
Before describing it, we need a few definitions.
For any vertex $u$ in a graph $G$, let $\delta_G(u)$ denote the number of edges incident to $u$
(the vertex degree) and let $\Delta_G=\max_u \delta_G (u)$.
If $G$ is an instance of the bounded edge coloring problem, we let $G^k$ be the subgraph
of $G$ containing edges with bounds $\geq k$ and we let $G_k$ be the subgraph
containing edges with bounds $\leq k$.
\begin{figure}[t]
\centerline{\includegraphics[width=2.5in]{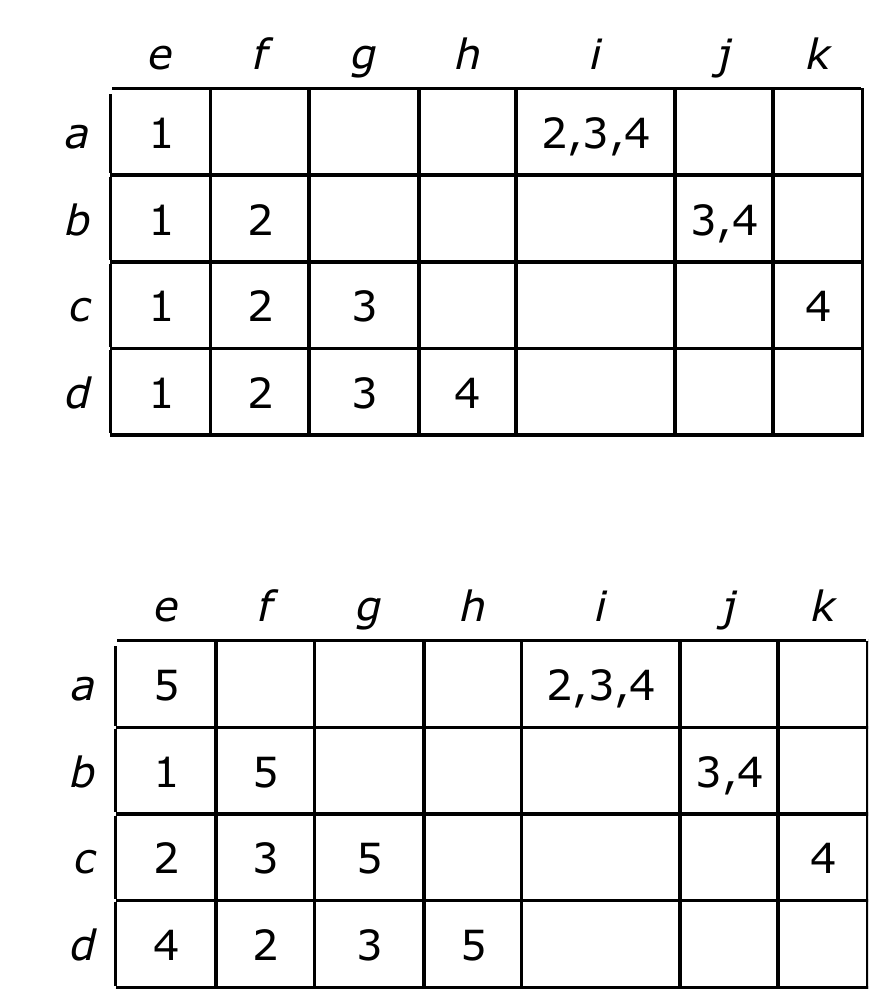}}
\caption{$D$ colors are not always enough (top table shows edge bounds, bottom shows colors in an optimum coloring)}
\label{example3}
\end{figure}

Observe that if a vertex $u$ has $d$ edges with bounds $\geq c$, then some edge
incident to $u$ must be assigned a color $\geq c+d-1$.
Consequently, for any $k$, the number of colors needed to color $G^k$ is at least
$k+\Delta_{G^k}-1$. The degree bound for $G$ is denoted $D(G)$ and defined by
$D(G)=\max_k k+\Delta_{G^k}-1$.
For the graph in Figure~\ref{example2}, $D=5$ and this graph can be colored using colors
$1,\ldots ,5$. 
Some graphs require more than $D$ colors.
Figure~\ref{example3} shows a graph (in the tablular format)
with $D=4$ that requires five colors
(an optimal coloring is shown in the second table).

\begin{figure}[t]
\centerline{\includegraphics[width=4.5in]{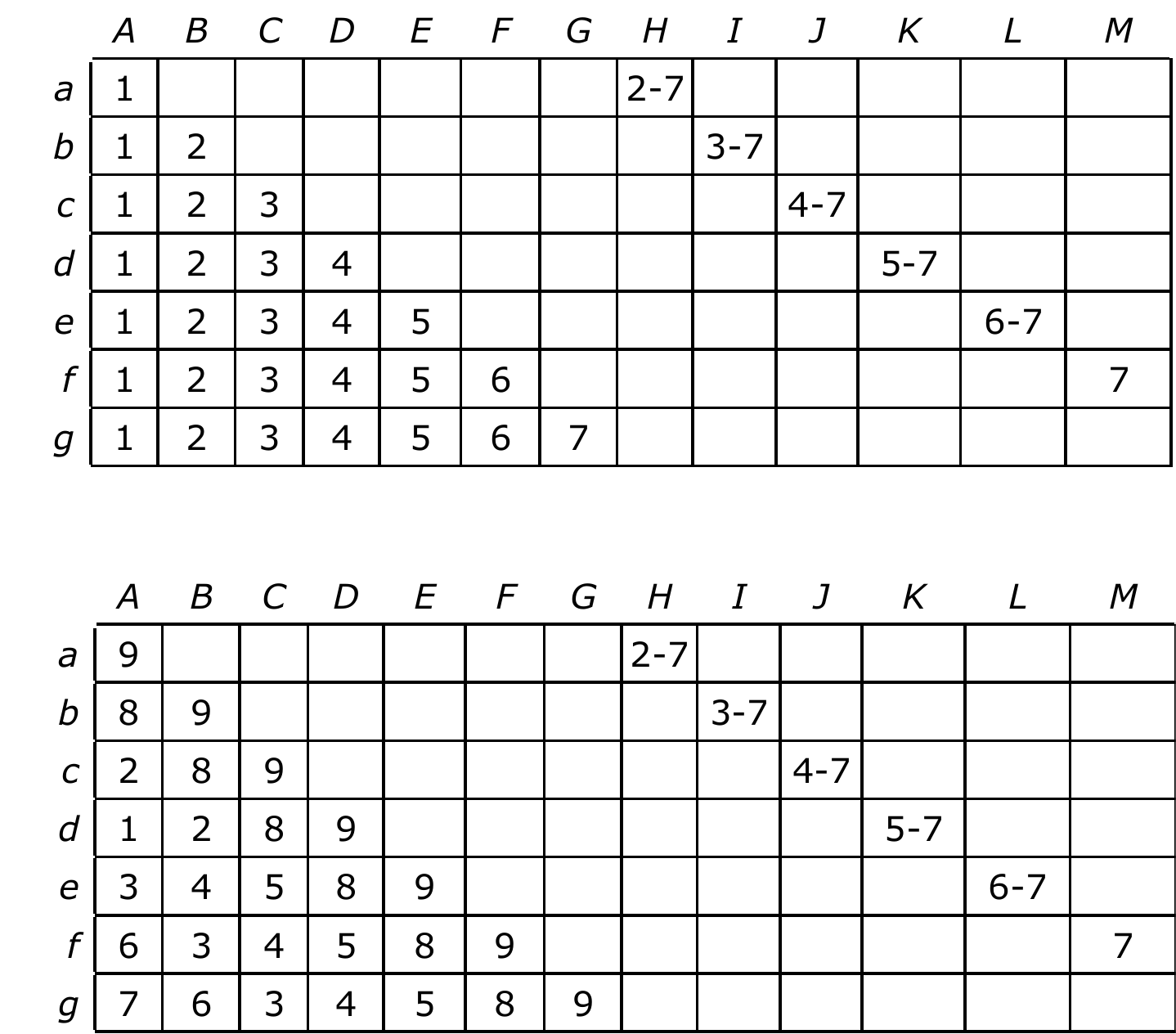}}
\caption{Graph $B_7$ and a valid edge coloring using colors $1,\ldots,9$}
\label{example4}
\end{figure}

The graph in Figure~\ref{example3} is actually a special case of a class of
graphs that require substantially more than $D$ colors. The graph $B_n$ has
inputs $u_1,\ldots,u_n$ and  outputs $v_1,\ldots,v_{2n-1}$.
For $1\leq i\leq n$, $1\leq j \leq i$, there is an edge $(u_i,v_j)$ with bound $j$,
and for each $1\leq i < n$, $i< j \leq n$, there is an edge $(u_i,v_{n+i})$
with bound $j$. Figure~\ref{example4} shows the case of $B_7$, along with
a coloring using nine colors. Note that in general, $D_{B_n}=n$.

Our second lower bound is obtained by computing a sequence of matchings.
Observe that for any integer $k$, the edges of $G$ that are colored $k$ must
form a matching in $G_k$.
If we let $m_k$ be the number of edges in a maximum size matching of $G_k$,
then $m_1+m_2+\cdots+m_k$ is an upper bound on the number of edges that
can be assigned colors in $1,\ldots,k$. If this sum is less than the number of edges in $G$,
then $G$ requires more than $k$ colors. So, if we let $M(G)$ be the smallest integer $k$
for which $m_1+m_2+\cdots+m_k$ is greater than or equal to the number of edges in $G$,
then $M(G)$ is a lower bound on the number of colors required to color $G$.
We refer to $M$ as the {\sl matching bound}. For the graph $B_4$ in Figure~\ref{example3},
the sequence of matching sizes is $1, 3, 4, 4, 4$ and consequently, $M=5$.
For the graph $B_7$ in Figure~\ref{example4},
the sequence of matching sizes is $1, 3, 5, 7, 7, 7, 7,7,7$ 
and $M=9$. It's not difficult to show that
in general, the matching bound for $B_n$ is $\lceil 5n/4 \rceil$.

Now, we turn to our third lower bound.
Let $G=(V,E)$ be a bipartite graph with edge bounds $b(e)$
and let $c$ be a valid coloring of $G$ using colors $1,\ldots,C$.
Let $H_k$ be the subgraph of $G$ that
is colored using colors $1,\ldots,k$  and let
$J_k$ be the subgraph of $G$ that is colored using colors $k+1,\ldots,C$.
Note that $H_k$ is a subgraph of $G_k$, $D_{H_k} \leq k$ and $\Delta_{J_k} \leq C-k$.
So, for any $k$, it is possible to split $G$ into two subgraphs that have
these properties.
Now, let $C^\prime$ be an integer smaller than $C$ and
note that if there is some $k$, for which we cannot split $G$ into
subgraphs $H_k$ and $J_k$ with $H_k$ a subgraph of $G_k$,
$D_{H_k} \leq k$ and $\Delta_{J_k} \leq C^\prime-k$, then
$G$ cannot be colored using only colors $1,\ldots,C^\prime$.
This leads to a lower bound on the number of colors needed to color a given graph.

To make this bound useful, we need an efficient way to partition $G$
into subgraphs $H_k$ and $J_k$ for given integers $k$ and $C$. This can be done
by solving a network flow problem. Let $F_{k,C}$ be a {\sl flow graph} that includes
a {\sl source vertex} $s$, a {\sl sink vertex} $t$ and a chain of vertices for each
vertex in $G_k$.
Specifically, for each input $u$ in $G_k$,
$F_{k,C}$ contains a chain consisting of vertices $u_i$ for each $i\in \{1,\ldots,k\}$
and edges $(u_i,u_{i+1})$ with {\sl capacity} $k-i$.
There is also an edge from $s$ to $u_1$ with capacity $k$.
For each output $v$ in $G_k$,
$F_{k,C}$ contains a chain consisting of vertices $v_i$ for each $i\in \{1,\ldots,k\}$
and edges $(v_{i+1},v_{i})$ with {\sl capacity} $k-i$.
There is also an edge from $v_1$ to $t$ with capacity $k$.
For each edge $(u,v)$ in $G_k$ with bound $i$, $F$ contains an edge $(u_i,v_i)$
of capacity 1.
We refer to this last set of edges as the {\sl core edges} of $F_{k,C}$.
Observe that the subset of the core edges that have positive capacity in any integer flow 
on $F_{k,C}$ correspond to a subgraph of $G_k$ that has a degree bound $D$ that is no larger than $k$.
To complete the construction of $F_{k,C}$, we specify {\sl minimum flow requirements} for the
edges incident to $s$ and $t$. In particular, for input $u$ of $G$,
the edge $(s,u_1)$ is assigned a minimum flow of $\min \{0,\delta_G(u)-(C - k)\}$.
Similarly, for output $v$ of $G$,
the edge $(v_1,t)$ is assigned a minimum flow of $\min \{0,\delta_G(v)-(C - k)\}$.
Given an integer flow on $F_k$ that satisfies the minimum flow requirements,
we define $H_k$ to consist of those edges in $G$ that correspond to
core edges that have positive flow. We define $J_k$ to include the remaining edges in $G$.
It is straightforward to show that $H_k$ is a subgraph of $G_k$, 
$D_{H_k} \leq k$ and $\Delta_{J_k} \leq C-k$.
If there is no integer flow on $F_k$ that satisfies the minimum flow requirements,
then $G$ cannot be colored using only colors $1,\ldots,C$.

\begin{figure}[t]
\centerline{\includegraphics[width=5in]{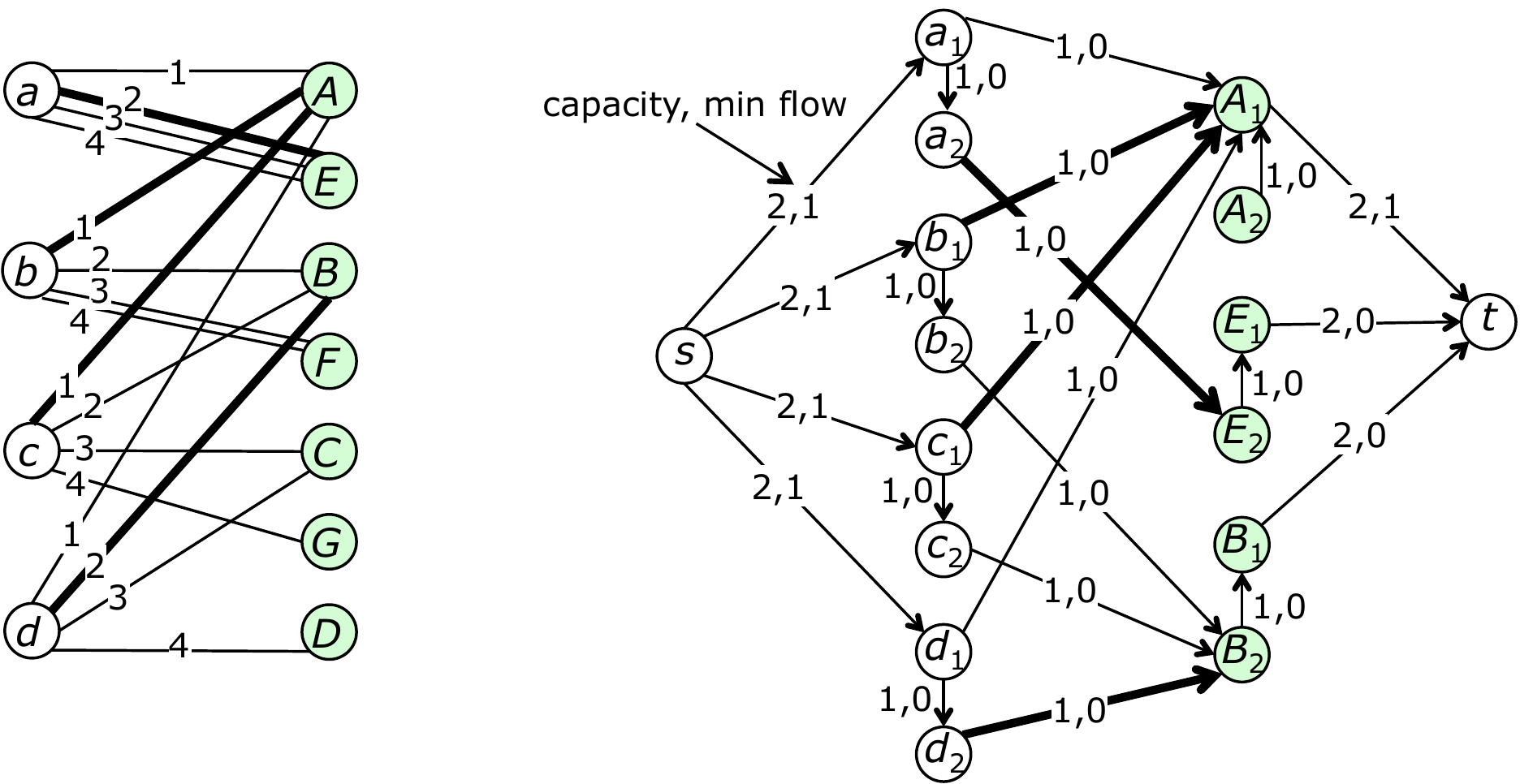}}
\caption{Graph $B_4$ and corresponding flow graph $F_{2,5}$ }
\label{flowBound}
\end{figure}

Figure~\ref{flowBound} shows the graph $B_4$ and the corresponding
flow graph $F_{2,5}$. There is an integer flow for $F_{2,5}$ that
uses the edges that are emphasized in ``{\bf bold}'' and satisfies
all the minimum flow requirements. The edges of $B_4$ that correspond to the
bold core edges are also emphasized in bold. 
These edges define the graph $H_2$, while the remaining
edges define $J_2$. Note that $D_{H_2}=2$ and $\Delta_{J_2}=3$.

The {\sl flow bound} for $G$ is denoted by $P_G$ and is defined as the smallest
value of $C$ for which $F_{k,C}$ has a flow that satisfies the minimum flow requirements,
for all values of $k\in[1,\bmax]$, where $\bmax$ is the 
largest edge bound in $G$.
The flow bound for $B_8$ is 11, while the matching bound is 10.
This gap increases for larger graphs. For example, $B_{64}$ has a
matching bound of 80 and a flow bound of 83, while
$B_{256}$ has a matching bound of 320 and a flow bound of 331.

We close this section by describing a general method for coloring
graphs $B_n$ using colors $1,\ldots,n+\lceil (n-1)/3 \rceil$.
For $B_{64}$ the largest color is 85, for $B_{256}$ the largest color is 341.
\begin{figure}[t]
\centerline{\includegraphics[width=4in]{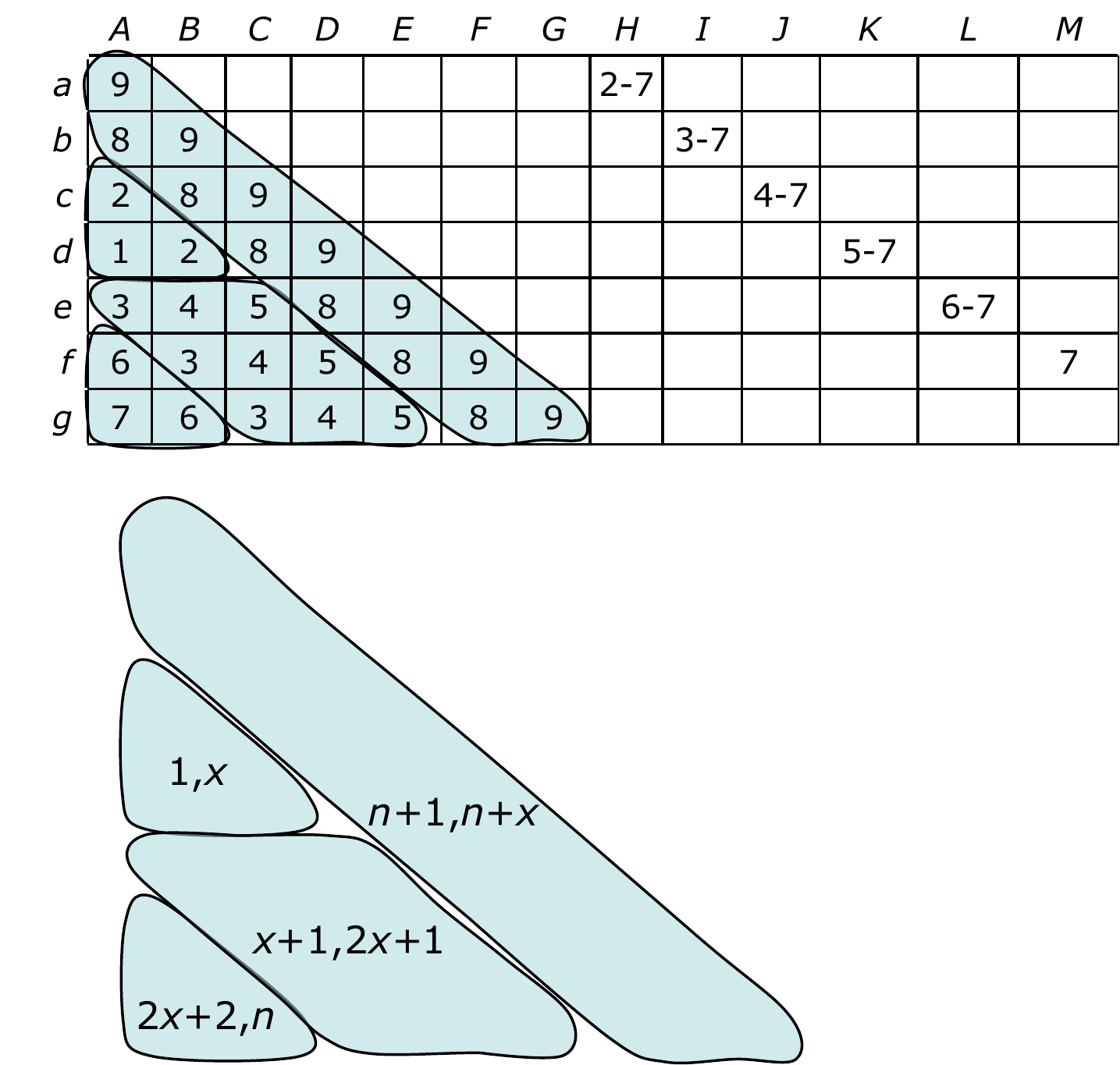}}
\caption{General method for coloring $B_n$}
\label{colorB}
\end{figure}
Figure~\ref{colorB} illustrates the method for coloring $B_n$.
The top part of the figure shows the coloring given earlier for $B_7$,
with several regions of the table highlighted. Note that each of the
four highlighted regions uses a distinct set of colors and each color used is
repeated along a diagonal within the region. The edges incident to
the outputs $H,\ldots,M$ are assigned colors equal to their bounds.
The bottom part of the figure shows how the colors of the edges incident to the
first $n$ outputs are assigned in the general case, using a parameter $x=\lceil (n-1)/3 \rceil$.
Again, we divide the table definining the edges into four regions and assign
disjoint sets of colors to those regions. Within each region, colors are used repeatedly
along diagonals. The edges incident to the last $n-1$ outputs are assigned colors
equal to their bounds. The choice of $x$ ensures that the assignment of colors
to edges yields a legal coloring of the graph that respects all the edge bounds.

\section{Approximate algorithms}

In this section, we describe several approximate algorithms for the bounded
edge coloring problem and analyze their worst-case performance.
Let $G=(V,E)$ be a graph with edge bounds $b(e)$ and let $C^\ast (G)$ be
the largest color value used by an optimal solution.

We start with a very simple method that produces solutions with a
maximum color $\leq 2 C^\ast$.
Let $\bmax$ be the largest color bound used by an instance of the bounded
edge coloring problem and let $\Delta$ be the maximum vertex degree.
Using any method for the ordinary edge coloring problem, we can color
the edges using colors $\bmax,\ldots,\bmax+\Delta -1$ and since
$C^\ast\geq \max\{\bmax,\Delta\}$,
the largest color used by this solution is $<2C^\ast$.

We can get a better approximation by first spliting $G$ into two subgraphs
and coloring each of the subgraphs separately.
Consider an optimal coloring of $G$ and let $H_k$ be the subgraph defined by
the edges with colors $\leq k$. Note that $H_k$ is a subgraph of $G_k$ and
that $\Delta_{H_k}\leq k$. If we let $J_k$ be the subgraph of $G$ defined by
the edges with colors $>k$, then $\Delta_{J_k}\leq C^\ast - k$.
So an optimal coloring must be divisible into a pair of subgraphs that
satisfy these inequalities. Note that we can color $H_k$ using colors
$k,\ldots,2k-1$ using any algorithm for the ordinary edge coloring problem.
We can also color $J_k$ using colors $\bmax,\ldots,\bmax+C^\ast-(k+1)$.
So long as $2k-1<\bmax$, these two sets of colors do not overlap meaning that
we can color the entire graph using colors $k,\ldots,\bmax+C^\ast-(k+1)$.
If we let $k=\lfloor \bmax/2 \rfloor$, the largest color is
$\leq (\bmax/2)+C^\ast\leq (3/2)C^\ast$.

In order to construct an approximation algorithm based on this observation,
we need a way to split $G$ into subgraphs $H_k$ and $J_k$. This can be done
by solving the same flow problem that was used in the flow lower bound computation.
Alternatively, we can simplify the flow problem by collapsing each of the input-side
chains $u_1,\ldots,u_k$
into a single vertex $u$ and each of the output-side chains $v_1,\ldots,v_k$
into a single vertex $v$.
Thus, we can efficiently color any graph $G$ using colors 
$\leq (3/2)C^\ast$. We'll refer to this as the {\sl splitting method}.

Our next algorithm is a simple greedy algorithm that repeats the following step until
all edges are colored.
\ipar
Select an edge $e=(u,v)$ and let $c$ be the smallest color that is at least as large as $b(e)$
and is not yet in use at both $u$ and $v$.
\rapi
We select edges that are incident to a vertex of maximum degree in the uncolored subgraph.
While we have no worst-case-performance
bound for this algorithm, in pactice it out-performs the splitting method, as we will see in the
next section.

Now, we consider an algorithm based on the classical augmenting path algorithm for the
ordinary edge coloring problem. For the bounded edge coloring problem, a path $p$
is an $ij$-{\sl augmenting path} if its edges alternate in color between $i$ and $j$, it cannot
be extended any further at either of its endpoints and every edge in the path has a bound
that is $\leq \min\{i,j\}$. The {\sl augmenting path algorithm} for the bounded edge coloring problem
colors the edges by repeatingly selecting an edge $e=(u,v)$ and then applying the first case from
the following list that applies.
\begin{itemize}
\item
 If there is some {\sl eligible color} that is unused
at both endpoints, color $e$ with one such color.
\item
If there are eligible colors $i$ and $j$, where $i$ is available at $u$ and $j$ is available at $v$,
and there is an $ij$-augmenting path starting at $v$,
then reverse the colors of the edges on the path and and let $c(e)=i$.
\item
If there are eligible colors $i$ and $j$, where $i$ is available at $u$ and $j$ is available at $v$,
and there is a $ji$-augmenting path starting at $u$,
then reverse the colors of the edges on the path and let $c(e)=j$.
\item
Allocate a new eligible color and use it to color $e$.
\end{itemize}
Initially, colors $1,\ldots,\bmax$ are eligible. New colors are allocated sequentially, as needed.
We select edges that are incident to a vertex of maximum degree in the uncolored subgraph.
When selecting colors, we give preference to colors with smaller values.
In the first case, we select the smallest eligible color that is unused at both endpoints.
In the second and third cases, we select color pairs $i$ and $j$ that minimize $\max\{i,j\}$.

Since the augmenting path algorithm for the ordinary edge-coloring problem
uses $\Delta$ colors, this version colors uses no color larger than $\bmax+\Delta -1$.
We can also use it to color the subgraphs in the splitting method to obtain an algorithm 
that uses no color larger $(\bmax/2)+C^\ast$.

Next, we consider algorithms based on constructing a series of matchings.
The first such algorithm starts by initializing $k=1$ then repeating 
the following step so long as there are uncolored edges.
\ipar
Find a maximum size matching on the uncolored edges in $G_k$,
assign color $k$ to all edges in the matching, then increment $k$.
\rapi
We refer to this as the {\sl maximum size matching} algorithm.
We can improve it by selecting matchings that maximize the number of matched
vertices that have maximum degree in the uncolored subgraph.
This can be done using an algorithm described in~\cite{tu15}.
This version of the matching algorithm is called the {\sl maximum degree matching} algorithm.
Reference~\cite{tu15} also shows how to find maximum size matchings that 
maximize a general {\sl priority score} based on arbitrary integer priorities assigned
to the vertices. Our third matching algorithm uses this method.
Priorities are assigned in decreasing order of vertex degree in the uncolored subgraph.
So, vertices of maximum degree are assigned priority 1, those with the next largest degree
are assigned priority 2 and so forth. We refer to this version of the matching algorithm
as the {\sl priority matching} algorithm. The last two of the matching algorithms
use no color larger $\bmax+\Delta -1$. Also, like the augmenting path algorithm,
they can be used with the splitting method to obtain an algorithm 
that uses no color larger $(\bmax/2)+C^\ast$.

\section{Experimental Evaluation}

\begin{figure}[t]
\centerline{\includegraphics[width=3.5in]{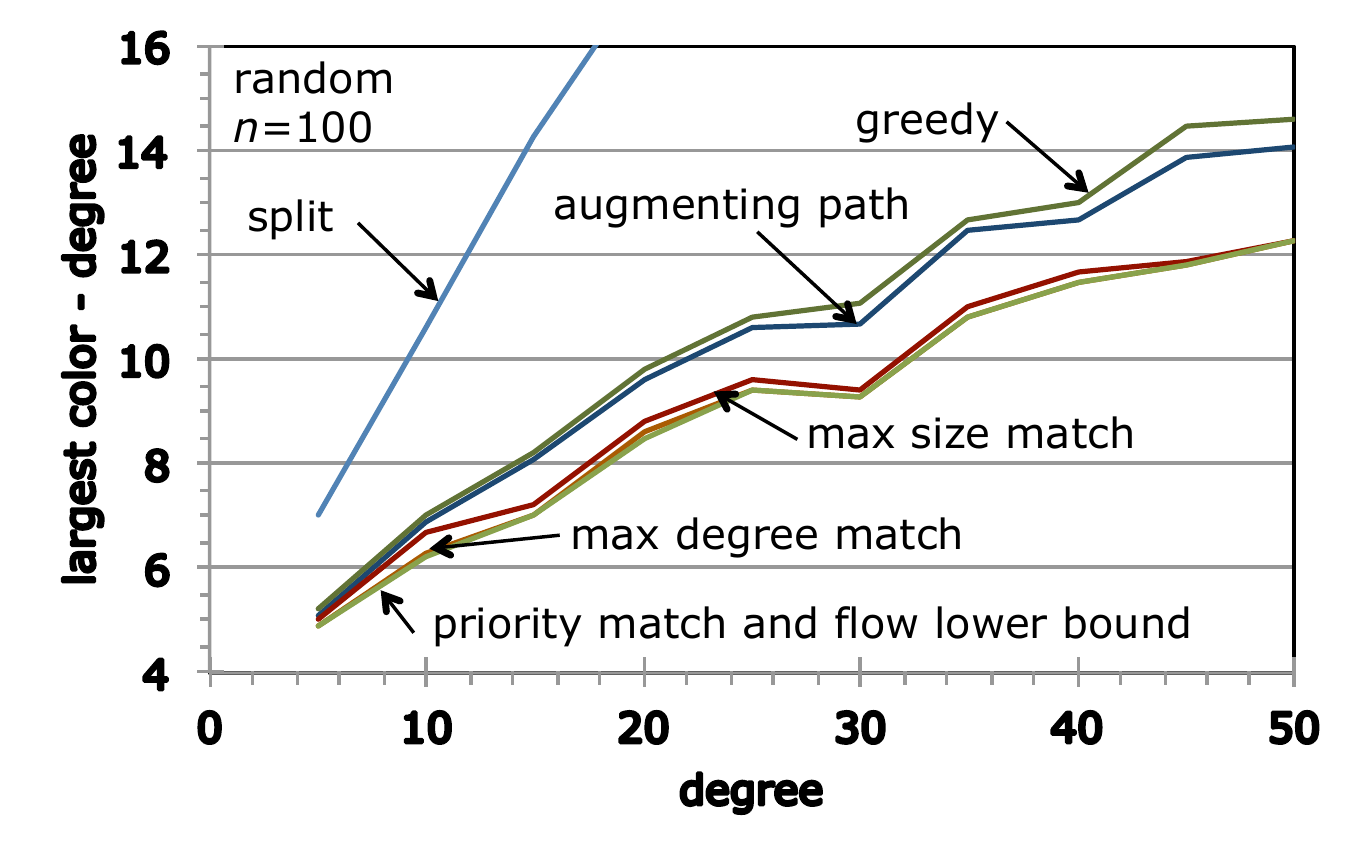}}
\caption{Performance results for random instances}
\label{randomchart}
\end{figure}

In this section, we evaluate the performance of the algorithms introduced in the
last section, experimentally. We start, by evaluating the performance using random graphs
that are generated using the following procedure.
\begin{itemize}
\item
Generate a random regular bipartite graph, with $n$ inputs, $n$ outputs and all
vertices having degree $\Delta$.
\item
At each input, assign the incident edges a unique random color in $1,\ldots,\bmax$,
where $\bmax \geq \Delta$
\end{itemize}
This produces a random problem instance that satisfies the unique input bounds condition.
Figure~\ref{randomchart} shows results for graphs on 100 vertices where the vertex degree
is varied from 5 to 50 and $\bmax=\Delta+3$. The $y$-axis displays the difference between
the maximum color used and the vertex degree. Each data point shows the average from ten
random problem instances. Error bars have been omitted for clarity, but the relative error
was generally less than 2\%.

The lowest curve shows the results for the flow lower bound and the priority matching algorithm
(for these graphs, the priority matching algorithm always produced results equal to the bound).
The maximum degree matching algorithm was almost identical to the priority matching algorithm,
but did occasionally exhibit small differences.
The maximum size matching algorithm was not quite as good as the other two.
The greedy and augmenting path algorithms also produced perfectly respectable results,
generally exceeding the largest color used by the priority matching algorithm by less than 5\%.

\begin{figure}[t]
\centerline{\includegraphics[width=3.5in]{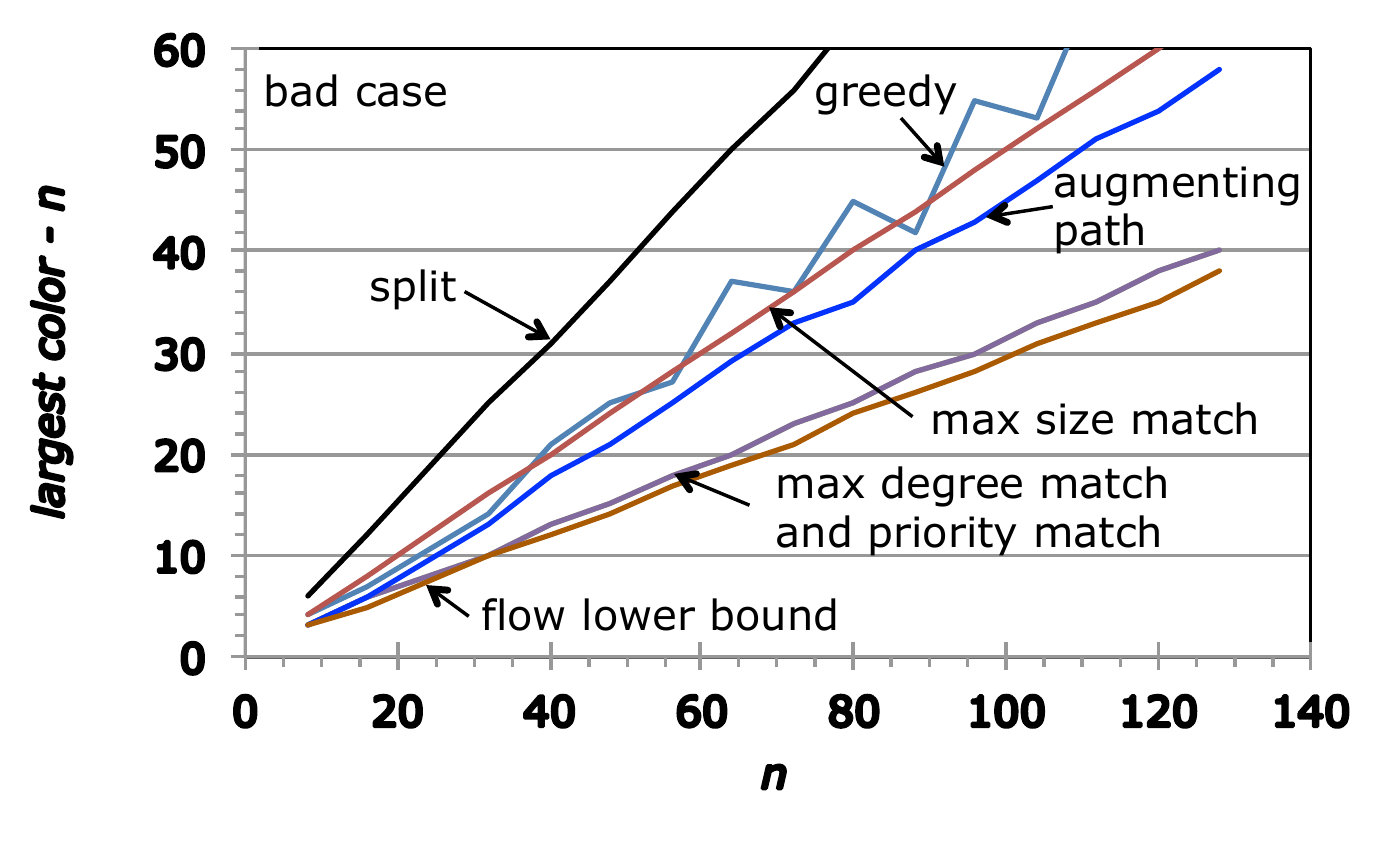}}
\caption{Performance results for $B_n$}
\label{badcasechart}
\end{figure}

Figure~\ref{badcasechart} shows performance results for the graphs $B_n$.
Here the max degree and priority match algorithms generally use just 2 or 3 more
colors than the number given by the flow lower bound. The maximum size matching
algorithm performs less well in this case, actually under-performing the augmenting path
algorithm.

The algorithms described here are all available as part of an open-source library
of graph algorithms and data structures~\cite{tu15a}.

\section{Closing Remarks}

The bounded edge-coloring problem is a natural generalization of the ordinary
edge-coloring problem and interesting in its own right, independent of its application to
the crossbar scheduling problem. There is a considerable gap between the best
worst-case performance ratio achieved by our algorithms and the experimental
performance measurements. Closing that gap is the main open problem to be addressed.

It's worth noting that most of our algorithms can be applied to general graphs,
as well as bipartite graphs. It would be interesting to understand how they perform
in this context. Unfortunately, the flow lower bound cannot be applied to general
graphs, although the degree bound and matching bound can be.

While our results do not apply directly to the online version of the crossbar scheduling
problem, there is potential for extending them to make them more applicable.
For example, one can model systems in which the crossbars operate at faster speeds
than the inputs and outputs by restricting the values allowed as edge bounds.
This can be used to derive a lower bound on the ``speedup ratio'' needed to
match the performance of an ideal output-queued switch.

\end{document}